\documentclass[reprint,superscriptaddress,showpacs,preprintnumbers,amsmath,amssymb,onecolumn,notitlepage]{revtex4-1}
\usepackage[T1]{fontenc}
\usepackage[polish]{babel}
\usepackage[utf8]{inputenc}
\usepackage{epsfig}
\usepackage{color}
\usepackage{graphicx}
\usepackage{subfigure}
\usepackage{amssymb,amsfonts,amsmath}
\usepackage{tikz}

\usepackage{graphicx}
\usepackage{bm}
\usepackage{hyperref}
\begin{document}

\title{Lagrangian statistics of dense emulsions}

\author{Ivan Girotto}
\affiliation{The Abdus Salam, International Centre for Theoretical Physics, Trieste, Italy}
\affiliation{Department of Applied Physics, Eindhoven University of Technology, Eindhoven, The Netherlands}
\author{Andrea Scagliarini}
\affiliation{Institute for Applied Mathematics ``Mauro Picone'' (IAC), Italian National Research Council (CNR), Rome, Italy}
\author{Roberto Benzi}
\affiliation{Department of Physics and INFN, University of Tor Vergata, Rome, Italy}
\author{Federico Toschi}
\affiliation{Department of Applied Physics, Eindhoven University of Technology, Eindhoven, The Netherlands}
\affiliation{Institute for Applied Mathematics ``Mauro Picone'' (IAC), Italian National Research Council (CNR), Rome, Italy}

\begin{abstract}

\noindent The dynamics of dense
stabilized emulsions presents a rich phenomenology including
chaotic emulsification, non-Newtonian rheology and ageing dynamics at rest.
Macroscopic rheology results from the complex droplet microdynamics and, in turn,
droplet dynamics is influenced by macroscopic flows via the competing action of hydrodynamic and interfacial stresses,
giving rise to a complex tangle of elastoplastic effects, diffusion, breakups and coalescence events.
This tight multiscale coupling, together with the daunting challenge
of experimentally investigating droplets  under flow, hindered the
understanding of dense emulsions dynamics.
We present results from 3d numerical simulations of dense
stabilised emulsions, resolving the shape and dynamics of individual droplets,
along with the macroscopic flows. We investigate droplet dispersion statistics, measuring
probability density functions (PDF) of droplet displacements and velocities,
changing the concentration, in the stirred and 
ageing regimes. We provide the first measurements ever, in concentrated emulsions, of the relative droplet-droplet
separations PDF and of the droplet acceleration
PDF, which becomes strongly non-Gaussian as the volume
fraction is increased above the jamming point. Cooperative effects, arising when droplets are
in contact, are argued to be responsible of the anomalous superdiffusive behaviour of
the mean square displacement and of the pair separation at long times, in both the stirred and
in the ageing regimes. This superdiffusive behaviour is reflected in a non-Gaussian pair separation
PDF, whose analytical form is investigated, in the ageing regime, by means of theoretical
arguments.
This work paves the way to developing a connection between Lagrangian
dynamics and rheology in dense stabilized emulsions.  
 
\end{abstract}

\maketitle

\section{Introduction}\label{sec:intro}
Understanding the dynamics of dense suspensions of soft, athermal particles such as emulsions, foams or gels is crucial for
many natural and industrial processes~\citep{lar99,clementsfood,coussot}.
A key question concerns the connection between mechanisms occurring at the 
microstructure level with the macroscopic flow and rheological properties in these
systems~\citep{pit13,dollet14,bonn17,dijksman19}.  
For instance, irreversible topological rearrangements, corresponding to local yielding 
events, are known to be directly related to the inhomogeneous fluidisation of 
soft glassy materials~\citep{Goyon2008,BocquetPRL2009,Bouzid2015,Dollet2015,Fei2020}.
A clear comprehension of the relevant processes and time-scales characterizing the microdynamics relies on
tracking single material mesoconstituents
(droplets, bubbles, etc)~\citep{SquiresMason,Durian1991,MasonPRL1997,Durian1995,CipellettiFD2003,RuzickaPRL2004,CerbinoPRL2008}.
Highly packed emulsions/foams are typically characterized in simple flows (oscillatory Couette, Poiseuille, etc), 
or even at rest, in the ageing regime~\citep{MasonPRL1995,CipellettiFD2003,Ramos2001,Cipelletti2005,Li2019,GiavazziJPCM2021}.
Lagrangian studies of dispersions in complex, high
Reynolds number flows, on the other hand, are widely represented in the literature,
but in extremely diluted conditions~\citep{Toschi2009,Brandt2022}. The investigation of the microdynamics of
concentrated systems in complex flows, of relevance, e.g., for emulsification processes~\citep{Vankova1,Vankova2,Vankova3},
is, in fact, a formidable task due to the need to cope, at the same time, with the interface dynamics two-way-coupled
to the hydrodynamics and with the droplet/bubble tracking.
This is what we address here, namely the statistical Lagrangian
dynamics of droplets in dense emulsions subjected to chaotic flows. We remark that this is the first investigation of this kind.
We employ a mesoscopic numerical method recently developed
to simulate the hydrodynamics of immiscible fluid mixtures, stabilized against full phase separation.
In a previous contribution, we showed how, by means of a suitable combination of chaotic stirring and injection
of the disperse phase, it is possible to prepare a three dimensional dense emulsion, that was then rheologically characterized, evidencing its
yield stress character~\citep{girotto2022}.
In the present paper, we discuss and employ a tracking algorithm for the trajectories of individual droplets
to investigate the droplet dynamics, in both semi-diluted and highly concentrated conditions,
under stirring and during ageing.
We study the statistics of droplet velocities and accelerations, focusing on the detection of
non-Gaussian signatures and how they are related to the nature of
droplet-droplet interactions. 
We discuss single and pair droplet dispersion,
showing that at high volume fractions a superdiffusive behaviour is observed in both
the stirred and ageing regimes. For pair dispersion in the ageing regime we propose theoretical models that show good
agreement with the measurements. Let us underline that measurements of the droplet acceleration PDFs and of the droplet pair
dispersion in densely packed emulsions have not been addressed before. 
Remarkably, our results suggest that both non-Gaussian statistics and superdiffusion emerge as soon
as the volume fraction exceeds a value comparable with that of random close packing of spheres, to be considered a proxy of the jamming
point. Therefore, this phenomenology is likely to be ascribable to cooperative effects resulting from the complex elastoplastic dynamics of
the emulsion.
The paper is organized as follows. In section \ref{sec:methods} we present the numerical method and we provide an extensive introduction to the tracking algorithm.
The main results are reported in section \ref{sec:results}, organized in subsections relative to the stirred and ageing regimes.
Conclusions and perspectives are drawn in section \ref{sec:conclusions}.

\section{Methods}\label{sec:methods}

\subsection{Multicomponent emulsion modeling}\label{sec:numerical_models}
Our numerical model is based on a three-dimensional (3D) two-component
lattice Boltzmann method~\citep{BENZI1992145}
in the Shan-Chen formulation~\citep{PhysRevE.47.1815, PhysRevE.49.2941}.
The lattice Boltzmann equation for the discrete probability distribution functions, $f_{\sigma l}$, reads
(the time step is set to unity, $\Delta = 1$)
\begin{equation}
    f_{\sigma l}(\mathbf{x}+\mathbf{e}_l, t + 1) - f_{\sigma l}(\mathbf{x}, t) = 
    -\frac{1}{\tau_{\sigma}}\left(f_{\sigma l}(\mathbf{x}, t) - f_{\sigma l}^{\text{eq}}(\mathbf{x}, t) \right) + S^{\text{(tot)}}_{\sigma l}(\mathbf{x}, t)
\end{equation}
where the index $l$ runs over the discrete set of nineteen 3D lattice velocities ($D3Q19$ model) $\{\mathbf{e}_l\}$
($l=0,1,\dots,18$),
and $\sigma$ labels each of the two immiscible fluids,
conventionally indicated as $O$ and $W$ (for, e.g., 'oil' and 'water').
The equilibrium distribution function is given by 
the usual polynomial expansion of the Maxwell-Boltzmann distribution, valid in the limit of small fluid velocity, namely:
\begin{equation}
  f_{\sigma l}^{\text{eq}}(\mathbf{x}, t) = \rho_{\sigma}\omega_l\left(1 + \frac{\mathbf{e}_l\cdot\mathbf{u}}{c^2_s} + 
    \frac{(\mathbf{e}_l\cdot\mathbf{u})^2}{2c_s^4} - \frac{\mathbf{u}\cdot\mathbf{u}}{2c^2_s} \right)
\end{equation}
with $\omega_l$ being the usual set of suitably chosen weights so to maximise the algebraic
degree of precision in the computation of the hydrodynamic fields  
and $c_s=1/\sqrt{3}$ a characteristic molecular velocity (a constant of the model).
The hydrodynamical fields (densities and total momentum) can be computed out of the lattice probability density functions $f_{\sigma l}$ as
$\rho_{\sigma} = \sum_l f_{\sigma l}$ and
$\rho \mathbf{u} = \sum_{\sigma l} f_{\sigma l}\mathbf{e}_l$
(where $\rho = \sum_{\sigma} \rho_{\sigma}$ is the total fluid density).
The source term $S^{\text{(tot)}}_{\sigma l} = \omega_l \mathbf{e}_l \cdot \mathbf{F}^{\text{(tot)}}_{\sigma}/c_s^2$ stems from all the forces (internal and external) acting in the system,
$\mathbf{F}_{\sigma} = \mathbf{F}_{\sigma} + \mathbf{F}_{\sigma}^{\text{(ext)}}$. In particular, $\mathbf{F}_{\sigma}$, incorporates the two kinds of lattice interaction forces,
$\mathbf{F}_{\sigma} = \mathbf{F}_{\sigma}^{\text{(r)}} + \mathbf{F}_{\sigma}^{\text{(f)}}$ and $\mathbf{F}_{\sigma}^{\text{(r)}}$ is the standard Shan-Chen inter-species repulsion
of amplitude, which is responsible of phase separation, and reads $G_{\text{OW}}>0$:
\begin{equation}\label{eq:LBint1}
    \mathbf{F}_{\sigma}^{(\text{r})}(\mathbf{x}, t) = 
    -G_{\text{OW}}\rho_{\sigma}(\mathbf{x}, t)\sum_{l,\sigma\neq\bar{\sigma}}\omega_l \rho_{\bar{\sigma}}(\mathbf{x}+\mathbf{e}_l, t)\mathbf{e}_l.
\end{equation}
The second term, $\mathbf{F}_{\sigma}^{\text{(f)}}$, consists of a short range intra-species attraction, involving
only nearest-neighbouring sites ($\mathcal{I}_1$), and a long range self repulsion,
extended up to next-to-nearest-neighbours ($\mathcal{I}_2$)~\citep{doi:10.1063/1.3216105}, namely:
\begin{align}\label{eq:LBint2}
\begin{split}
  \mathbf{F}_{\sigma}^{\text{(f)}}(\mathbf{r}, t) = &-G_{\sigma\sigma,1}\rho_{\sigma}(\mathbf{x}, t)\sum_{l\in \mathcal{I}_1}\omega_l
  \rho_{\sigma}(\mathbf{x}+\mathbf{e}_l, t)\mathbf{e}_l  \\
  &-G_{\sigma\sigma,2}\rho_{\sigma}(\mathbf{x}, t)
  \sum_{l\in \mathcal{I}_1 \bigcup \mathcal{I}_2}p_l\rho_{\sigma}(\mathbf{x}+\mathbf{e}_l, t)\mathbf{e}_l,
\end{split}
\end{align}
where $G_{\text{OO},1}, G_{\text{WW},1} < 0$ and $G_{\text{OO},2}, G_{\text{WW},2} > 0$ and $p_l$ are the weights of
the $D3Q125$ model.
This type of repulsive interaction $G_{\sigma\sigma,2}$
represents a mesoscopic phenomenological modelling of surfactants and provides a mechanism of stabilisation against coalescence of close-to-contact droplets
(the superscript 'f' stands in fact for 'frustration'),
promoting the emergence of a positive disjoining pressure, $\Pi>0$, within the liquid film separating the approaching
interfaces~\citep{doi:10.1063/1.3216105, Benzi2010, C4SM00348A}.\\
The large-scale forcing needed to generate the chaotic stirring that mixes the two fluids enters into the model through $\mathbf{F}_{\sigma}^{\text{(ext)}}$,
which takes the following form:
\begin{equation}
  F^{\text{(ext)}}_{\sigma i}(\mathbf{x}, t) = A \rho_{\sigma}\sum_{j \neq i}\left[\sin(k_j r_j+\Phi_{k}^{(j)}(t))\right],
\label{eq:turbo}
\end{equation}
where $i,j=1,2,3$, $A$ is the forcing amplitude, $k_j$ are the wavevector components, and the sum is limited to $k^{2}=k_{1}^{2}+k_{2}^{2}+k_{3}^{2}\leq2$. 
The phases $\Phi_{k}^{(j)}$ are evolved in time according to independent Ornstein-Uhlenbeck processes with the same relaxation times $T=L/u_{\text{rms}}$,
where $L$ is the cubic box edge and $u_{\text{rms}}$ is a typical large-scale
velocity~\citep{Biferale_2011,doi:10.1063/1.4719144}.

\subsection{Droplet tracking}\label{sec:tracking}
We present now the tracking algorithm and discuss its implementation. The algorithm combines (i) the process of identification
of all droplets constituting the emulsion at two different and consecutive time steps (hereafter called {\it labeling}),
with (ii) a stage describing the kinematics of each droplet (the actual {\it tracking}).\\
In the labeling step, individual droplets, defined as connected clusters of lattice points such that the local density
exceeds a prescribed threshold (equal to $\rho_O^{(\text{max})}-\rho_O^{(\text{min})}$),
are identified by means of the Hoshen-Kopleman algorithm~\citep{PhysRevB.14.3438,FRIJTERS201592}.
This approach echoes what is known in the image processing jargon as
Connected Component Labeling~\citep{HE201725}; similar techniques have been recently applied to multiphase fluid
dynamics in Volume of Fluid simulations~\citep{Herrmann2010,HENDRICKSON2020104373}.\\  
The tracking is based on the computation of the probability to obtain volume transfer among droplets in space and time
as described in~\citep{10.1016/j.jcp.2022.111560,GAO2021103523,CHAN2021110156}.
This is performed as follows. 
Let us suppose that in the domain, at time $t_1$, there are $N_1$ droplets and at a later dump
$t_2 = t_1 + \Delta t$ there are $N_2$ droplets.
%We set the density field to $0$ value on all lattice points where there are no droplets, and
%to a value ($\geq{1}$) on all lattice sites inside a droplet.
We define a set of droplet indicator fields $\rho_k(x,t)$ (with $k$ running over the all droplets),
which are equal to $k$ if, at time $t$, $x$ is inside the $k$-droplet, 
and are equal to $0$ elsewhere. 
In the following, the ``state'' representing the droplet will be denoted in the ket notation
(reminiscent of quantum mechanics states)
as $|k,t\rangle$ (the state is assumed to be normalized by square root of the droplet volume, $\sqrt{V_k}$),
such that the transition probability for a droplet $k_1$ at a time $t_1$ to end up in a droplet
$k_2$ at a time $t_2$ is given by the bra-ket expression:
\begin{equation}\label{eq:tracking}
  P_{k_1\rightarrow k_2}=\langle k_2,t_2|k_1,t_1\rangle  = \frac{1}{V} \int \rho_{k_2}(\mathbf{x},t_2)\rho_{k_1}(\mathbf{x},t_1)d\mathbf{x}, \quad V=\sqrt{V_{k_1} V_{k_2}}
\end{equation}
This transition probability is equal to $1$ in the case droplets $k_1$ and $k_2$ perfectly overlap and it is zero if they do not overlap at all.
A high $P$ value gives us, therefore, the confidence in having re-identified the same droplet at two different time steps.
What happens if a droplet is not deforming and just translating with uniform velocity $\mathbf{v}$?
We expect that the probability will decrease due to an imperfect overlap between the droplet $k$ at time $t$ and the same droplet $k$, displaced
by $\mathbf{v}\Delta t$ at time $t+\Delta t$, where $\mathbf{v}$ is the average velocity of all grid points included into a droplet. Therefore, we expect that the maximal correlation will occur for:
\begin{equation}
\langle k,t+\Delta t|k,t\rangle={1\over V}\int \rho_k(\mathbf{x}+\mathbf{v}\Delta t,t)\rho_k(\mathbf{x},t+\Delta{t})d\mathbf{x}
\label{eq:tracking_2}
\end{equation}
Of course the amount of this effect is proportional to $\Delta t$. In order to reduce the effect of the translation
of the droplet $k$ at a given $\Delta t$
we implement a Kalman filter, evaluating the overlap against the predicted at the same
$t+\Delta t$ shifting the initial position at time $t$ forwards by $\mathbf{v}\Delta t$.
For all the data shown, the tracking is implemented with $\Delta t = 100$ lattice Boltzmann time steps.

\section{Results}\label{sec:results} 

\begin{figure}
\begin{center}
  \advance\leftskip-0.55cm
    \includegraphics[width=0.6\textwidth,angle=270]{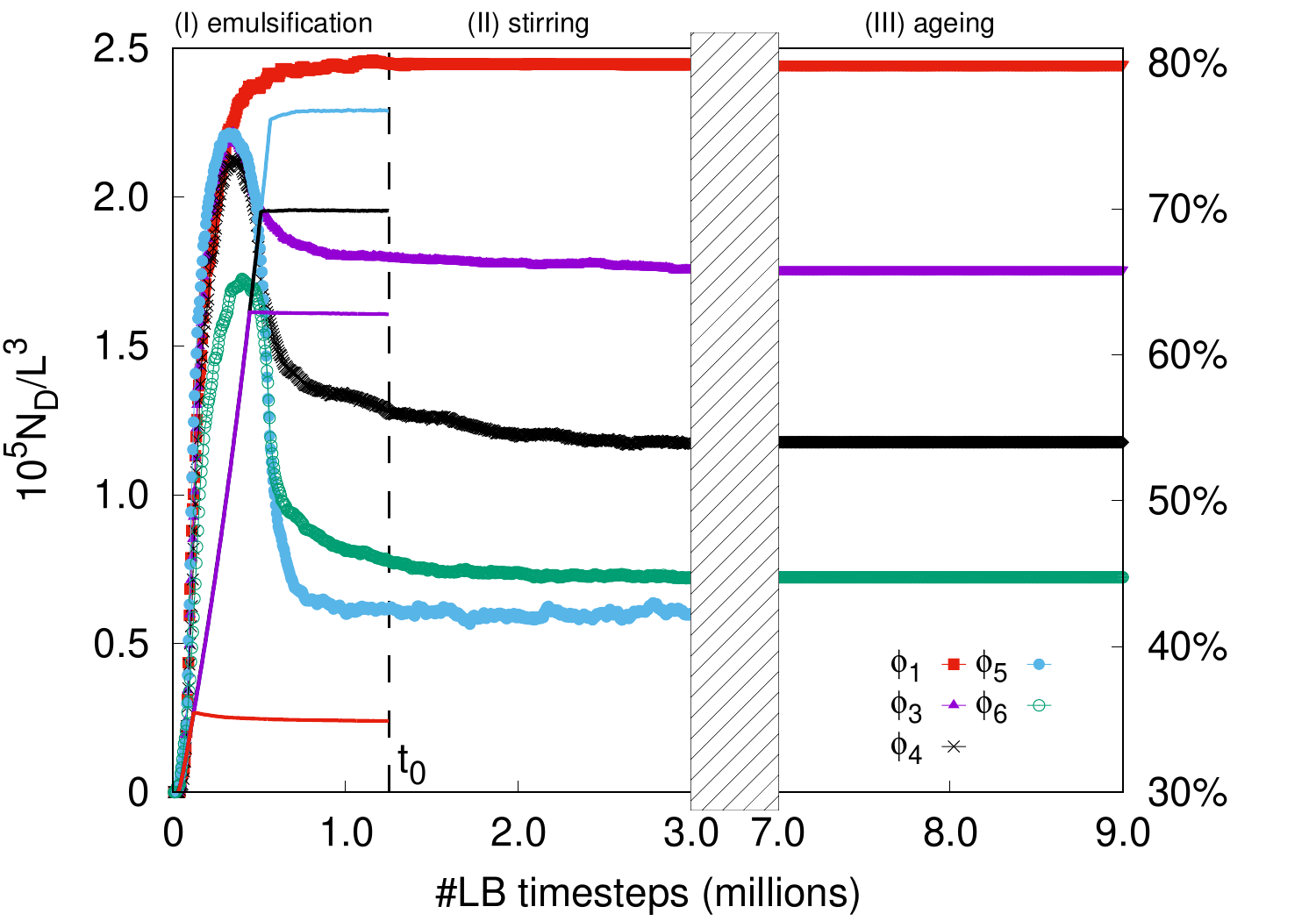}
    \caption{Number density of droplets, $N_D/L^3$, as a function of time for a set of four simulations, labelled with the
      corresponding target (steady state) volume fractions ($\phi$) of the dispersed phase
      (see Table~\ref{table:vol_frac_interface} for further details).
      The solid lines (color coded as for the droplet number density) indicate the time evolution of the volume fraction
      during emulsification.
      The vertical dashed line highlights the starting time of tracking, $t_0 = 1.25 \cdot 10^6$. 
      All simulations are stirred with the same forcing amplitude parameter 
      ($A=4.85\cdot 10^{-7}$, see Eq.~(\ref{eq:turbo})), except for the case indicated with hollow circles ($A=4.05 \cdot 10^{-7}$,
      run labelled as $\phi_6$ in Table~\ref{table:vol_frac_interface}). The relaxation phase $t \in [t_F, t_A^{(i)}]$ is omitted
      for the sake of clarity of visualisation.}
    \label{fig:events_n_drops}
\end{center}
\end{figure}
In this section we provide results aimed at characterizing the dynamics of individual droplets
(e.g., their velocities and accelerations, as well as absolute and relative dispersion), at changing the volume
fraction of the dispersed phase from $\phi=38\%$ to $\phi=77\%$. All simulations were performed on a cubic grid of side
$L=512$ lattice points, the kinematic viscosity was
$\nu=1/6$ for both components, (in lattice Boltzmann units; hereafter dimensional quantities will be all expressed
in such units) and the total density $\rho_f = 1.36$ (giving a dynamics viscosity $\eta = \rho_f \nu \approx 0.23$).
With reference to Fig.~\ref{fig:events_n_drops}, where we plot the temporal variation
of the number density of droplets $N_D/L^3$,
we give first a cursory description of the {/it emulsification} process, indicated as phase $(I)$ in the figure (further details can be found in~\citet{girotto2022}).
All simulations are run for a total of $9 \cdot 10^6$ time steps. 
Starting from an initial condition with $\phi=30\%$, where the two components are fully separated by a flat interface, 
the emulsion is created applying the large-scale stirring force, Eq.~(\ref{eq:turbo}), with magnitude $A=4.85\cdot 10^{-7}$,
while injecting the dispersed phase until the desired value is reached.
The duration of the injection phase, $t_{\text{inj}}$, depends, then, on the target volume fraction
(see Table~\ref{table:vol_frac_interface}). 
The forcing is applied up to $t_F=3 \cdot 10^6$. The evolution of the system is then monitored for further $6 \cdot 10^6$ time steps.
%(not shown in Fig.~\ref{fig:events_n_drops} since the number of droplets is essentially constant).
The tracking algorithm is activated at $t \geq t_0 = 1.25 \cdot 10^6$; for what we call, hereafter, {\it stirred regime} (phase (II) in Fig.~\ref{fig:events_n_drops})
we collect statistics over the interval $t \in [t_0,t_F]$ (which is statistically stationary, as it can be appreciated from the figure of $N_D(t)$, but also
looking at other observable, such as the mean square velocity), whereas, for the {\it ageing regime} (phase (III) in Fig.~\ref{fig:events_n_drops}),
we consider data in the interval $t \in [t_A^{(i)}, t_A^{(f)}]$, with $t_A^{(i)}= 7 \cdot 10^6$ and $t_A^{(f)}= 9 \cdot 10^6$,
(the intermediate relaxation phase, $t \in [t_F, t_A^{(i)}]$, is not shown in
Fig.~\ref{fig:events_n_drops}, for the sake of clarity of visualisation).
\begin{figure}
\centering
\subfigure[$\phi=38\%$]{
  \includegraphics[width=0.375\textwidth]{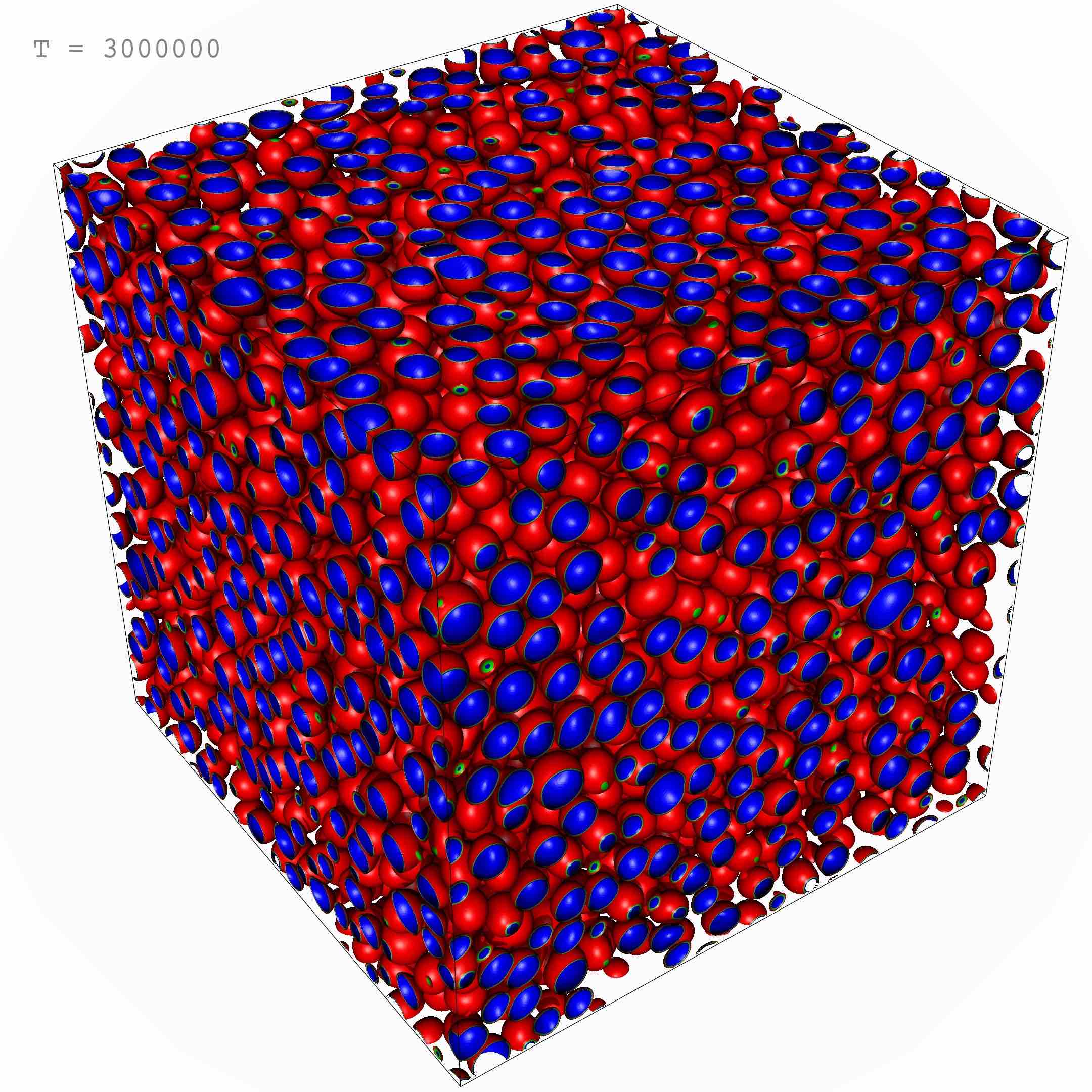}
  \label{fig:volume_fraction_interface_final-1}
}
\subfigure[$\phi=64\%$]{
  \includegraphics[width=0.375\textwidth]{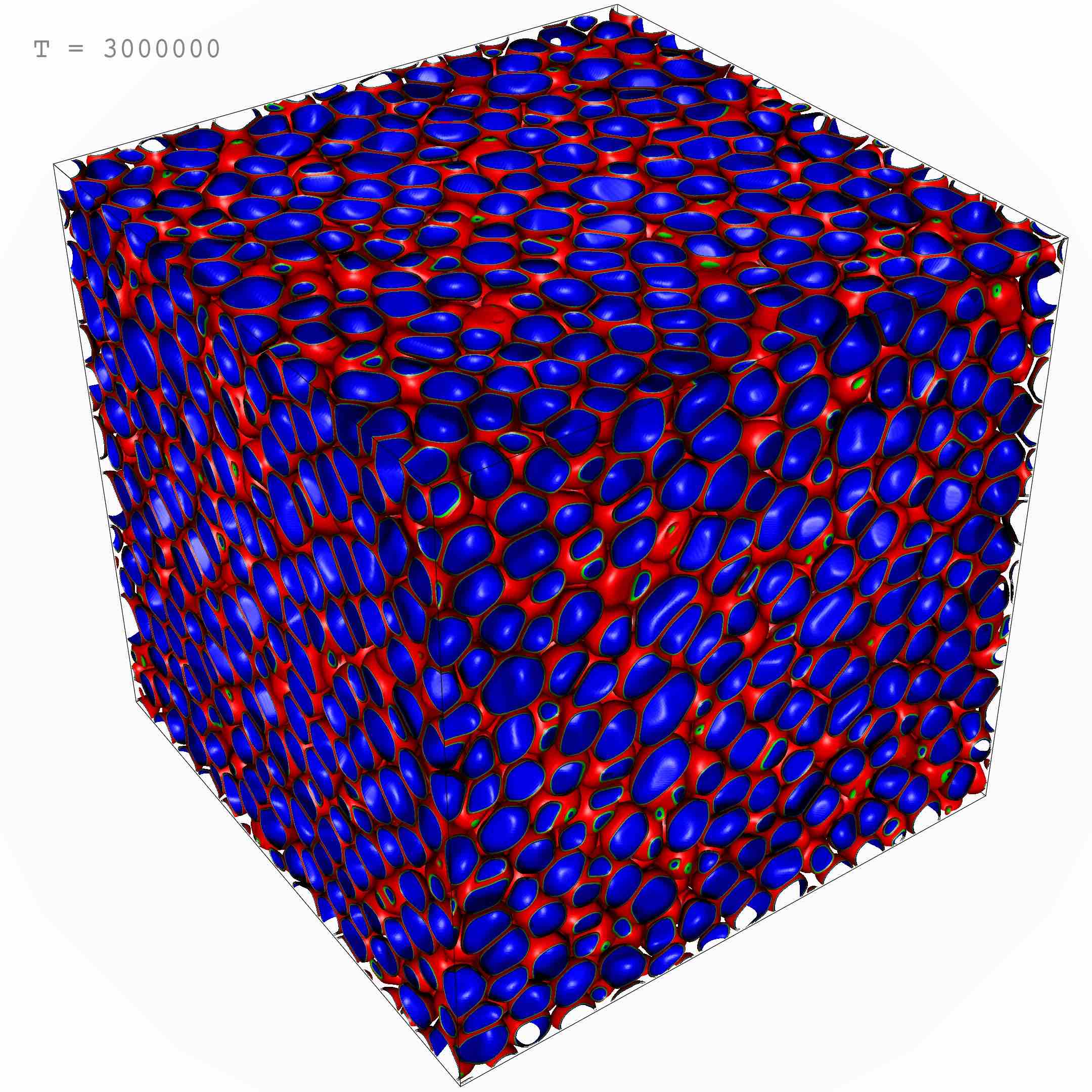}
  \label{fig:volume_fraction_interface_final-3}
}
\subfigure[$\phi=70\%$]{
  \includegraphics[width=0.375\textwidth]{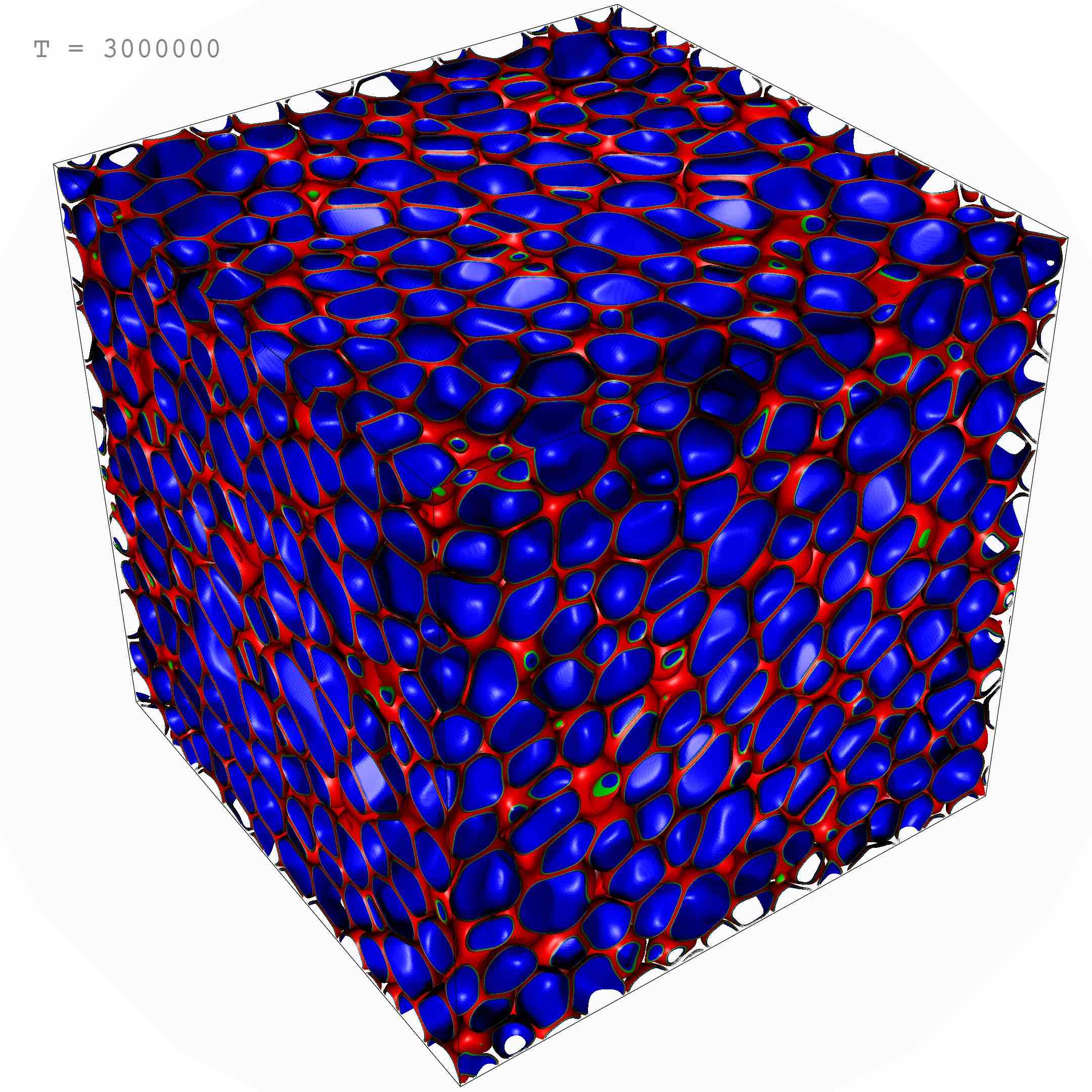}
  \label{fig:volume_fraction_interface_final-4}
}
\subfigure[$\phi=77\%$]{
  \includegraphics[width=0.375\textwidth]{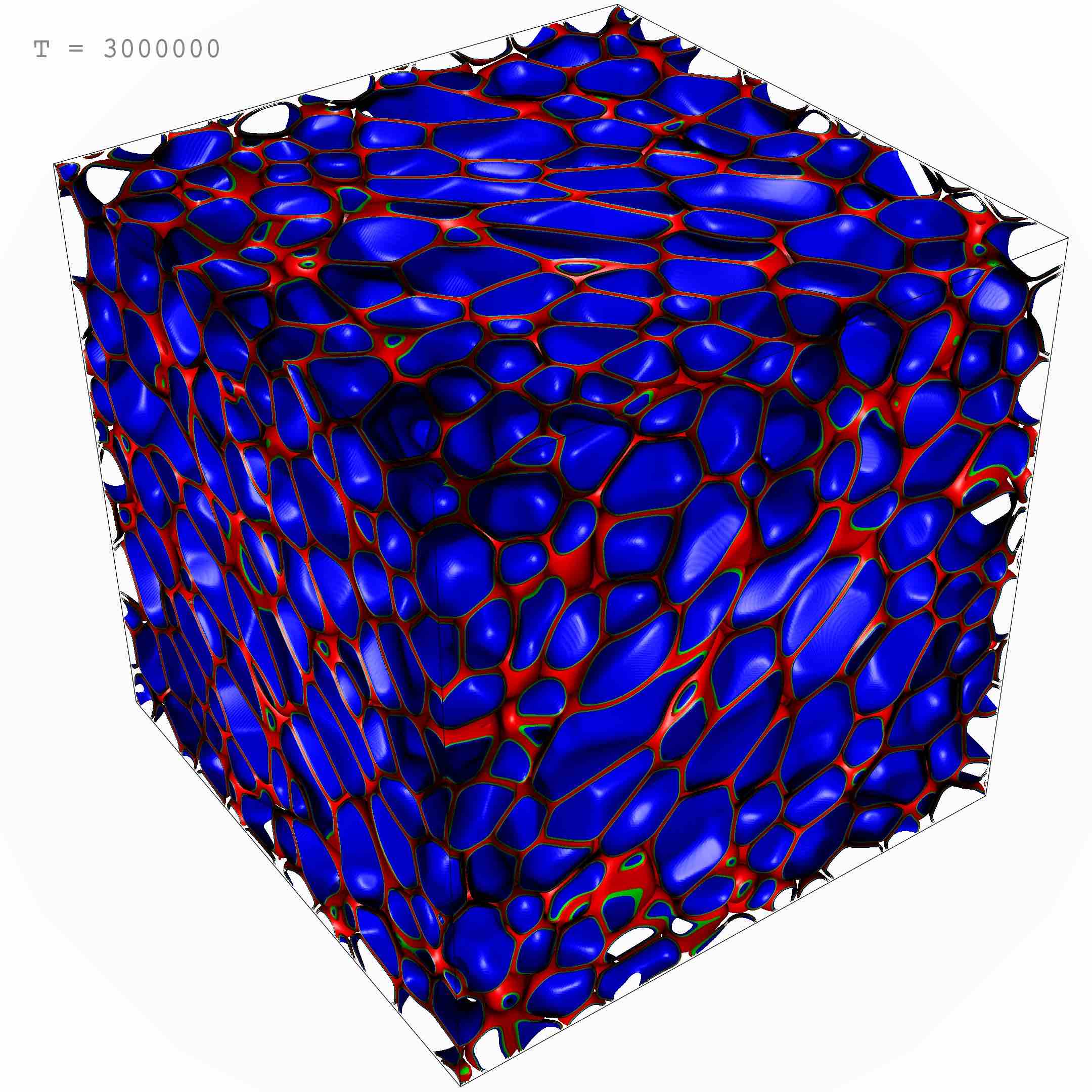}
  \label{fig:volume_fraction_interface_final-5}
}
\caption{Snapshots from the stirred regime showing the morphology of the emulsions.
  At incresing the volume fraction, $\phi$, a lower number of larger droplets is observed.
  Volume fractions below $\phi_c = 64 \%$ (a value compatible with the random close packing of spheres in 3D)
  are characterized by spherical droplets, whereas higher fractions ($\phi > \phi_c$) clearly show 
  highly deformed droplets. More quantitative details can be found in Table~\ref{table:vol_frac_interface}.}
\label{fig:interface_volume_fraction_final}
\end{figure}
In Fig.~\ref{fig:interface_volume_fraction_final} we show snapshots of the morphology of the emulsions at $t=t_F$, for different volume fractions.
Semi-diluted emulsions present a high number of small spherical droplets, whereas densely packed emulsions
are constituted of a smaller number of non-spherical larger droplets.
\begin{table*} 
  \centering
  \begin{tabular}{|c|c|c|c|c|c|c|c|c|c|c|c|c|c|}
 $\phi$ & $t_{\text{inj}}$ & $ \overline{N_D}^{(s)}$ & $U^{(s)}_{\text{rms}} $ & $a^{(s)}_{\text{rms}}$ & $T_L^{(s)}$ & $ \langle d \rangle $ & $d_{\text{rms}}$ & $t_G$ & $U^{(a)}_{\text{rms}} $ & $a^{(a)}_{\text{rms}}$  & $T_L^{(a)}$ \\\hline
    \toprule
 $\phi_1 = 38\%$ & $1.5 \cdot 10^5$ & $3300$ & $2.8 \cdot 10^{-2}$ & $4.6 \cdot 10^{-4}$ & $1.8 \cdot 10^4$ & $30$  & $4$ & $180$ & $2.0 \cdot 10^{-5}$ & $1.2 \cdot 10^{-7}$ & $2.6 \cdot 10^7$ \\
$\phi_2 = 49\%$ & $3 \cdot 10^5$ & $3100$ & $2.5 \cdot 10^{-2}$ & $3.6 \cdot 10^{-4}$ & $2.0 \cdot 10^4$ &  $34$  & $5$ & $97$ & $1.7 \cdot 10^{-5}$ & $7.7 \cdot 10^{-8}$ & $3.0 \cdot 10^7$ \\
$\phi_3 = 64\%$ & $4.6 \cdot 10^5$ & $2400$ & $2.0 \cdot 10^{-2}$ & $2.6 \cdot 10^{-4}$ & $2.6 \cdot 10^4$ & $40$  & $6$ & $80$ & $5.1 \cdot 10^{-5}$ & $2.6 \cdot 10^{-7}$ & $10^7$ \\
$\phi_4 = 70\%$ & $5.2 \cdot 10^5$ & $1600$ & $1.8 \cdot 10^{-2}$ & $2.2 \cdot 10^{-4}$ & $2.8 \cdot 10^4$ & $48$  & $8$ & $1.1$ & $6.0 \cdot 10^{-5}$ & $4.2 \cdot 10^{-7}$ & $8.5 \cdot 10^6$\\
$\phi_5 = 77\%$ & $5.8 \cdot 10^5$ &  $800$ & $1.7 \cdot 10^{-2}$ & $2.1 \cdot 10^{-4}$ & $3.0 \cdot 10^4$ & $60$  & $17$ & $0.1$ & $-$ & $-$ & $-$\\
$\phi_6 = 77\%$ & $5.8 \cdot 10^5$ &  $970$ & $-$ & $-$ & $-$ & $58$  & $12$ & $-$ & $7.0 \cdot 10^{-5}$ &  $6.7 \cdot 10^{-7}$ & $7.3 \cdot 10^6$\\
  \end{tabular}
  \caption{Relevant averaged quantities from the simulations at the different volume fractions $\phi$. The overline, $\overline{()}$,
    indicates time average, the brackets, $\langle () \rangle$, indicate an average over time and over the ensemble of droplets.
    The superscripts $(s,a)$ indicate that the averages have been taken over the statistically steady stirring regime ($t \in [t_0, t_F]$)
    and over the ageing regime ($t\in [7\cdot 10^6, 9\cdot 10^6]$), respectively.
    The various columns contain: $t_{\text{inj}}$, injection endtime; 
    $ \overline{N_D}^{(s)}$, mean droplet number (stirring);  $U^{(s)}_{\text{rms}}$, root mean square droplet velocity (stirring);
    $a^{(s)}_{\text{rms}}$, root mean square droplet acceleration (stirring); $T_L^{(s)}=L/U^{(s)}_{\text{rms}}$,
    large eddy turnover time (stirring); $ \langle d \rangle $, mean droplet diameter;
    $d_{\text{rms}}$, standard deviation of the droplet diameter; $t_G$, mean (dimensionless) droplet life time; $U^{(a)}_{\text{rms}} $, root mean square droplet velocity (ageing);
    $a^{(a)}_{\text{rms}}$, root mean square droplet acceleration (ageing); $T_L^{(a)}=L/U^{(a)}_{\text{rms}}$,
    large eddy turnover time (ageing).
    Numerical values of the simulations parameters are:
    kinematic viscosity, $\nu=1/6$; total fluid density, $\rho_f = 1.36$; surface tension, $\Gamma=0.0238$;
    forcing amplitude, $A=4.85\cdot 10^{-7}$ (except for the run $\phi_6$ for which $A=4.05 \cdot 10^{-7}$); simulation
    box side, $L=512$.}
  \label{table:vol_frac_interface}
\end{table*}
We report in Table~\ref{table:vol_frac_interface} the mean and root mean square (rms)
values of some relevant observables (averaged in space and in time over the stirred phase, $t\in [t_0, t_F]$, and over the ageing phase,
$t\in [7 \cdot 10^6, 9 \cdot 10^6]$, respectively),
for the various volume fractions considered. 
One can immediately notice that, at increasing the volume fraction, accelerations
and velocities rms decrease, implying a higher effective viscosity, while at the same time,
the trend of the root mean square droplet diameter, $d_{\text{rms}}$, shows an increase in polydispersity.\\
%It can be noticed as the initial interface fragmentation displays a rapid increase in the number of droplets, and therefore expected to be breakup dominated.
%On the other hand, the phase of injection needed to achieve the high-concentration is characterized by a rapid reduction of the number of droplets, so expected to be dominated
%by coalescences events.   
\begin{figure}
\begin{center}
  \advance\leftskip-0.55cm
  \includegraphics[width=0.7\textwidth]{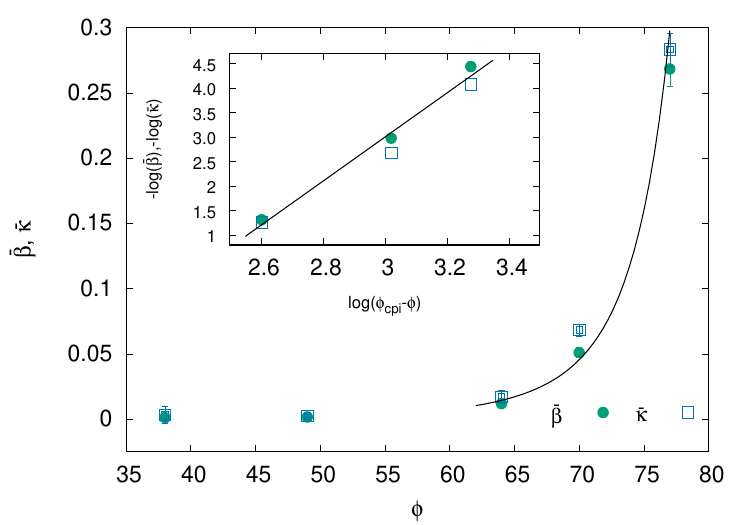}
  \caption{(Main panel). Mean breakup ($\overline{\beta}$) and coalescence ($\overline{\kappa}$) rates
    (averaged over stirred regime) as a function
    of the volume fraction concentration $\phi$. The rates equal each other,
    consistently with the dynamical equilibrium observed at steady state; both rates
    are very low ($\sim 10^{-3}$) and basically insensitive to the volume fraction below jamming ($\phi_c \approx 64\%$) and
    above it they increase steeply with $\phi$. The solid line represents the function
    $g(\phi) = \frac{C}{(\phi_{\text{cpi}} - \phi)^q}$, which is telling
    that the mean breakup and coalescence rates tend to diverge as the volume fraction approaches a critical value $\phi_{\text{cpi}}$,
    identifiable with that of catastrophic phase inversion (whence the subscript); fitting values of the parameters are
    $\phi_{\text{cpi}}=90.5\%$, $q=4.5$ and $C \approx 3.6 \cdot 10^4$.
    (Inset). Logarithms of the mean breakup and coalescence rates for $\phi \geq \phi_c$ as function of the
    logarithmic distance from $\phi_{\text{cpi}}$.}
   \label{fig:events_vol_frac}
\end{center}
\end{figure}
Fig.~\ref{fig:events_vol_frac} shows the rates of breakup ($\overline{\beta}$) and coalescence ($\overline{\kappa}$),
i.e. the number of events per unit time, averaged over the stirred regime
as a function of the volume fraction $\phi$ (in the inset we report $\beta(t)$
and $\kappa(t)$ as a function of time for $\phi = 70\%, 77\%$). In the steady state the system is in a dynamical
equilibrium with $N_D(t)$ essentially constant (see Fig.~\ref{fig:events_n_drops}), therefore breakup and coalescence
rates approximately balance each other, $\beta(t) \approx \kappa(t)$; moreover,
both mean rates are extremely low ($\sim 10^{-3}$) for
$\phi < \phi_c = 64\%$, i.e. below jamming,
and increase steeply with $\phi$ for $\phi>\phi_c$. Interestingly $\phi_c$ is in the expected range of
volume fractions for
random close packing of spheres in 3D~\citep{Torquato2010}.
The growth of $\overline{\beta}$ and $\overline{\kappa}$ above $\phi_c$ is particularly steep, suggesting a divergent behaviour
as the volume fraction approaches a ``critical'' value which can be arguably identified with the occurrence of the catastrophic
phase inversion, $\phi_{\text{cpi}}$. In particular, the divergence turns out to have a power-law character (as highlighted
in the inset where we plot the cologarithm of $\overline{\beta}$ and $\overline{\kappa}$ vs $\log(\phi_{\text{cpi}}-\phi)$):
the solid line depicts the fitting function $g(\phi)=\frac{C}{(\phi_{\text{cpi}}-\phi)^q}$, with fitted values
$\phi_{\text{cpi}}=90.5\%$ and $q=4.5$.\\
The number of breakup and coalescence events depends, of course, on the intensity of the hydrodynamic stresses involved.
To see how these quantities
can give a flavour of the stability of the emulsion to the applied forcing, let us introduce an average  
{\it droplet life time}, $t_G = \frac{\overline{N_D}U_{\text{rms}}}{\overline{\beta}L}$,
adimensionalized by the large eddy turnover time; from Table~\ref{table:vol_frac_interface}, we see that, below
jamming ($\phi \leq \phi_c$) $t_G \gg 1$, i.e. the droplets, on average, tend to preserve their integrity along the whole
simulation, whereas in the densely packed systems ($\phi > \phi_c$) $t_G$ abruptly drops (down to $t_G \approx 0.1$ for
the largest volume fraction, $\phi=77\%$).

\subsection{Velocity and acceleration statistics: stirred regime}

\begin{figure}
\begin{center}
  \advance\leftskip-0.55cm
    \includegraphics[width=0.7\textwidth]{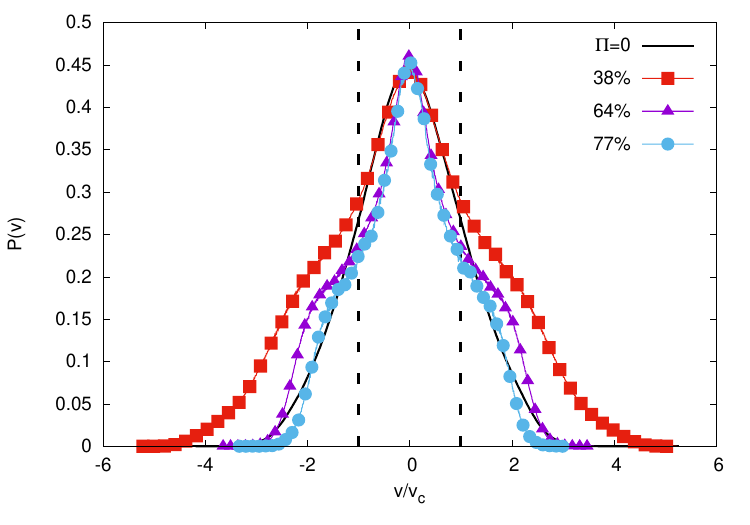}
     \caption{PDFs of the droplet velocities, for $\phi=38\%$, $\phi=64\%$ and $\phi=77\%$, computed over the
              forced steady state. The PDFs are normalized to have unitary area. 
              In both cases the curves show two marked inflection points,
              around $v \approx \pm v_c = 0.0175$,  which are, instead, suppressed when the disjoining pressure is deactivated, $\Pi=0$.}
    \label{fig:pdf_vel}
\end{center}
\end{figure}
In Fig.~\ref{fig:pdf_vel} we report the PDFs of the droplet velocities for volume fractions
$\phi=38\%$, $\phi=64\%$ and $\phi=77\%$. In both cases the PDFs show a bell shape, but in a range of intermediate
values of velocities $|v|$ they develop regions
with non-monotonic curvature, which are more
pronounced in the concentrated case. We ascribe such peculiar shape to the droplet-droplet interactions
(collisions and/or elastoplatic deformations). The characteristic velocity $v_c$ can be then estimated from
the balance of the elastic force and the Stokesian drag for a spherical droplet of diameter $d$, $F_S = 3\pi \eta d v_c$.
The elastic force acting on droplets squeezed against each other is due to the disjoining pressure $\Pi$ stabilising 
the inter-droplet thin film; at mechanical equilibrium the disjoining pressure equals the capillary pressure at the
curved droplet interface~\citep{DeryaguinChuraev}, therefore the force can be estimated as
the Laplace pressure times the cross sectional area,
$F_{\text{el}} \sim \frac{2\Gamma}{d}\pi \left(\frac{d}{2}\right)^2$, where $\Gamma$ is the surface tension. Letting
$F_{\text{el}} \sim F_S$ gives $v_c \sim \frac{\Gamma}{6\eta} = 0.0175$; from Fig.~\ref{fig:pdf_vel} we see
that, indeed, the inflections are located around $v\sim \pm v_c$.
To test our conjecture further, we have run a simulation setting the competing
lattice interactions responsible for the emergence of the disjoining pressure to zero
 ($G_{\text{OO},1}=G_{\text{WW},1}=G_{\text{OO},2}=G_{\text{WW},2}=0$ in Eq.~\ref{eq:LBint2}),
thus, effectively, we enforce $\Pi=0$. The resulting velocity PDF
is reported in Fig.~\ref{fig:pdf_vel} where we observe that, in fact, the inflectional regions disappear.\\
\begin{figure}
\begin{center}
  \advance\leftskip-0.55cm
    \includegraphics[width=0.7\textwidth]{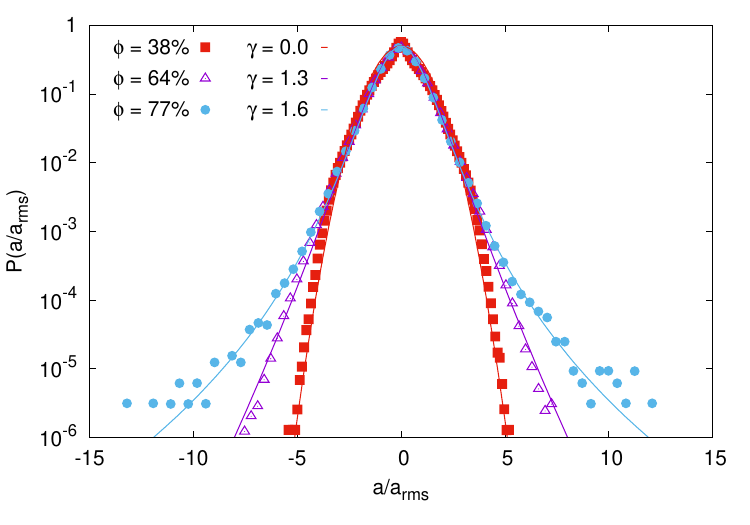}
    \caption{PDFs of the droplet accelerations, for $\phi=38\%, 64\%, 77\%$, computed over the,
      statistically steady, forced state. The PDFs are normalized to have unitary area. The values of
      $a_{\text{rms}}$ are given in Table~\ref{table:vol_frac_interface}.
      The solid lines are fits from Eq.~(\ref{eq:exp_dist}) with
            $\sigma=1$ and $\beta=0.35$, whereby the non-Gaussianity parameter $\gamma$ takes the values $\gamma=0$
            (corresponding to the Gaussian distribution) for $\phi=38\%$, $\gamma=1.32$ for $\phi=64\%$ and $\gamma=1.6$
             for $\phi=77\%$.}
    \label{fig:pdf_acc}
\end{center}
\end{figure}
The PDF of droplet accelerations, reported in Fig.~\ref{fig:pdf_acc}, is
Gaussian in the (semi)diluted emulsion ($\phi = 38\%$)  but, as the volume fraction is increased above 
$\phi_c = 64\%$ the PDFs tend to develop  {\it fat} (non-Gaussian) tails. 
A working fitting function is a stretched exponential of the type:
\begin{equation}
  P(\tilde{a})=C \exp\left(-\frac{\tilde{a}^{2}}{\left(1+\left(\frac{\beta |\tilde{a}|}{\sigma}\right)^{\gamma}\right)\sigma^{2}}\right),
  \label{eq:exp_dist}
\end{equation}
where $\tilde{a}=a/a_{\text{rms}}$.
The non-Gaussianity here, unlike turbulence, cannot be grounded on the complexity of the velocity field and of the
associated multifractal distribution of the turbulent energy dissipation~\citep{biferalePRL2004}.
We are not facing, in fact, a fully developed turbulent flow and, moreover, the non-Gaussian signatures become evident at increasing
the volume fraction above the jamming point $\phi_c$, where the effective viscosity is higher (and, hence, the effective Reynolds number is lower).
The origin of the non-Gaussianity is to be sought in the complex elastoplastic dynamics of the system, which is driven
by long-range correlated irreversible stress relaxation events.
Remarkably, in concentrated emulsions it has been shown that the spatial distribution of stress drops
in the system displays a multifractal character~\citep{KumarSciRep} which might be responsible
of the acceleration statistical properties at high volume fraction, in a formal analogy with the phenomenology of turbulence.
The curves corresponding to Eq.~(\ref{eq:exp_dist}) are reported in Fig.~\ref{fig:pdf_acc},
for various values of the parameter $\gamma$, which gauges the deviations from the Gaussian form,
and fixed $\beta=0.35$ (obtained from best fitting) and
$\sigma=1$, the standard deviation of the Gaussian limit, such that for $\gamma = 0$
the PDF reduces to the normal distribution $\propto e^{-x^2/2}$.
This is, indeed, the case for the lowest volume fraction, $\phi = 38\%$; the non-Gaussianity parameter $\gamma$ then
 increases monotonically with $\phi$ up to $\gamma = 1.6$ for $\phi = 77\%$.
%This is consistent with what found with the phenomenon of Rayleigh-B\'enard convection in dense emulsions
%at low Rayleigh number, where it was shown that, in spite of the system not being turbulent, the statistics
%of heat flux fluctuations was strongly non-Gaussian~\citep{pelusi2021}.

\subsection{Dispersion: stirred regime}
We focus now on the spatial dispersion of both single droplets as well as of droplet pairs.
The single droplet (or absolute) dispersion is defined in terms of 
the statistics of displacements, $\Delta \mathbf{X} = \mathbf{X}(t_0 + T) - \mathbf{X}(t_0)$,
where $\mathbf{X}(t)$ is the position of the droplet centre of mass at time $t$ ($t_0=1.25 \cdot 10^6$, let us recall, is the starting time of tracking).
\begin{figure}
\begin{center}
  \advance\leftskip-0.55cm
  \includegraphics[width=0.7\textwidth]{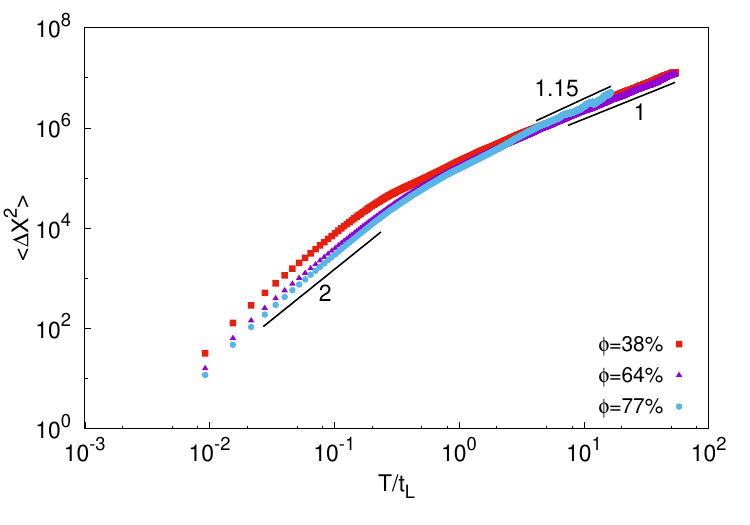}
  \caption{Mean square displacement (MSD), $\langle \Delta X^2 \rangle$ for $\phi=38\%, 64 \%$ and $77\%$,
    in the forced regime. The MSD goes initially as $T^2$, indicating a ballistic dynamics, followed by a diffusive
    growth, $T^{1/2}$, for $\phi < \phi_c$, whereas in the densely packed system ($\phi=77\%$) a superdiffusive behaviour
    $T^{\alpha}$, with $\alpha = 1.15$.}
  \label{fig:absolute_dispersion}
\end{center}
\end{figure}
In Fig.~\ref{fig:absolute_dispersion} we show the mean square displacement (MSD),
$\langle \Delta X^2 \rangle$ for volume fractions $\phi=38 \%, 64 \%$ and $77 \%$.
\begin{figure}
\begin{center}
  \advance\leftskip-0.55cm
  \includegraphics[width=0.7\textwidth]{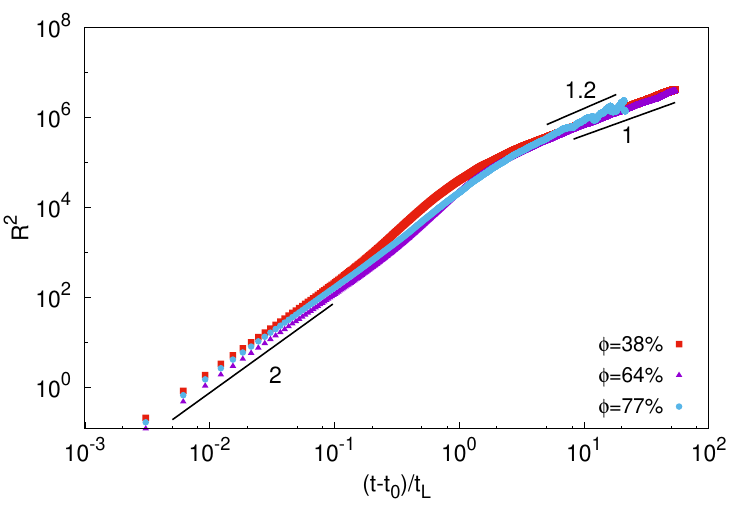}
  \caption{Mean square droplet pair separation $R^2(t) \equiv \langle |\mathbf{R}|^2 \rangle$
    as a function of time for volume fraction
    $\phi = 38\%, 64\%, 77\%$. The power laws are reported as solid lines, evidencing a slight deviation from the
    diffusive regime for the most concentrated case.}
  \label{fig:rel_dispersion_new}
\end{center}
\end{figure}
Values are reported in logarithmic scale, with the time normalised (hereafter) by a characteristic large scale time defined as $t_L = \sqrt{L/A} \approx 32500$ (which is independent of the volume fraction), where $A$ is the
amplitude of the applied forcing.
In the diluted case, the MSD (Fig.~\ref{fig:absolute_dispersion}) shows a crossover at around $T \sim 5t_d$ between an initial ballistic
motion, $\langle \Delta X^2 \rangle \sim T^2$,
and a diffusive behaviour at later times, $\langle \Delta X^2 \rangle \sim T$.
This is consistent with the typical Lagrangian dynamics of particles advected by
chaotic and turbulent flows~\citep{falkovich2001}; for intermediate times, though, we observe a transitional region, in correspondence,
approximately, of the crossover, where the curve presents an inflection point with locally super-ballistic slope. This is
an interestingly non-trivial behaviour that certainly deserves further investigation.
At increasing the volume fraction the long time growth becomes steeper, suggesting that a superdiffusive,
regime $T^{\alpha}$, with $\alpha = 1.15$, may occur.\\
A deeper insight on the small scale dynamics can be grasped by looking at the pair dispersion,
namely the statistics of separations
\begin{equation}\label{eq:R} 
\mathbf{R}_{ij}(t) = \mathbf{X}_i(t) - \mathbf{X}_j(t),
\end{equation}
at time $t$ for all pairs of droplets $i,j$ that are nearest neighbours (i.e. such that their corresponding cells
in a Voronoi tessellation of the centre of masses distribution are in contact) at $t=t_0$.
The observable in Eq.~(\ref{eq:R}) is, in fact, insensitive to contamination from mean homogeneous large scale flows,
if present.
In Fig.~ \ref{fig:rel_dispersion_new} we report the mean square value
$R^2(t) \equiv \langle |\mathbf{R}_{ij}|^2 \rangle_{\{ij\}}$ (where the average is over the initially neighbouring pairs)
as a function of time, for $\phi = 38\%, 64 \%$ and $77\%$.
%As the time goes by, the PDF becomes more and more spread and the maximum shift
%to increasingly larger values.
Analogously to the MSD, $R^2(t)$ grows in time and, after an initial ballistic transient,
it follows a $t^{\alpha}$ law, which is diffusive ($\alpha=1$) for concentrations below $\phi_c$ and
superdiffusive ($\alpha = 1.2$) for $\phi>\phi_c$.

\subsection{Velocity and acceleration statistics: ageing regime}

\begin{figure}
\begin{center}
  \advance\leftskip-0.55cm
  \includegraphics[scale=0.7]{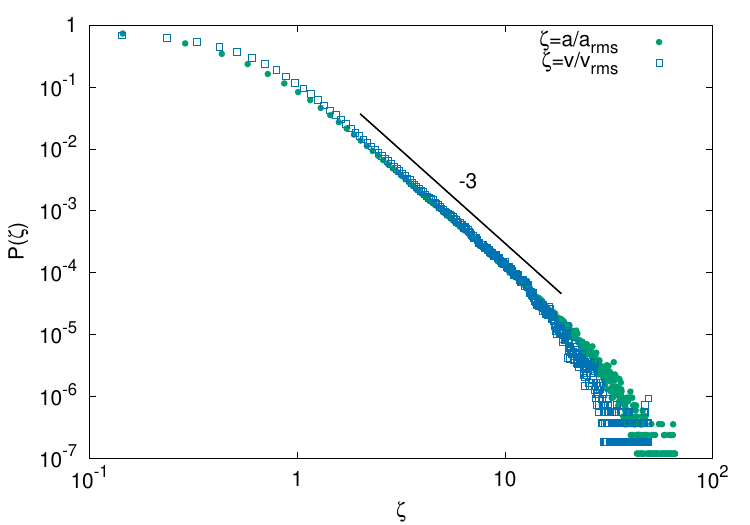}
  \caption{PDFs of the absolute values of droplet velocities and accelerations,
    rescaled by their own root mean square values
    (whose values are given in Table~\ref{table:vol_frac_interface}), in the ageing
    regime. Both PDFs show a power law, $P(\zeta) \sim \zeta^{-3}$, decay.}
\label{fig:pdfs_vel+acc}
\end{center}
\end{figure}
When the large scale forcing is switched off, in diluted conditions (below the close packing volume fraction),
the system relaxes
via a long transient where the kinetic energy decays to zero.
Instead, at high volume fraction (in the jammed phase), the emulsion is never completely at rest,
due to diffusion and droplets elasticity favouring the occurrence of plastic events,
local topological rearrangement of few droplets (i.e. during the ``ageing'' of the material).
Therefore, we consider here only the latter situation and focus on the case $\phi = 77\%$; hereafter, we present
data obtained with forcing amplitude $A=4.05 \cdot 10^{-7}$ (see the $\phi_6$ row in Table~\ref{table:vol_frac_interface}),
which yielded a larger number of droplets ($\sim 10^3$) in the steady state, such to improve the statistics.
The PDF of the droplet velocities is reported in Fig.~\ref{fig:pdfs_vel+acc}. Since there is no mean flow, the
PDF is an even function of its argument. We show, therefore, the distribution of the absolute values in logarithmic scale,
in order to highlight the power-law behaviour $v^{-3}$.
Interestingly, the PDF of acceleration also develops a power law tail $P(a) \sim a^{-3}$
(the PDFs for velocity and acceleration do, in fact, overlap, upon rescaling by the respective
standard deviations, see Fig.~\ref{fig:pdfs_vel+acc}), reflecting
the fact that, when stirring is switched off, the high effective viscosity overdamps the dynamics,
thus enslaving the acceleration to the velocity (by Stokesian drag), $a \sim u/\tau_s$
(assuming Stokes time equal for all droplets, which is reasonable given the very low spread of size
distribution in the ageing regime~\citep{girotto2022}).\\

\subsection{Dispersion: ageing regime} 

\begin{figure}
\begin{center}
  \advance\leftskip-0.55cm
  \includegraphics[scale=0.7]{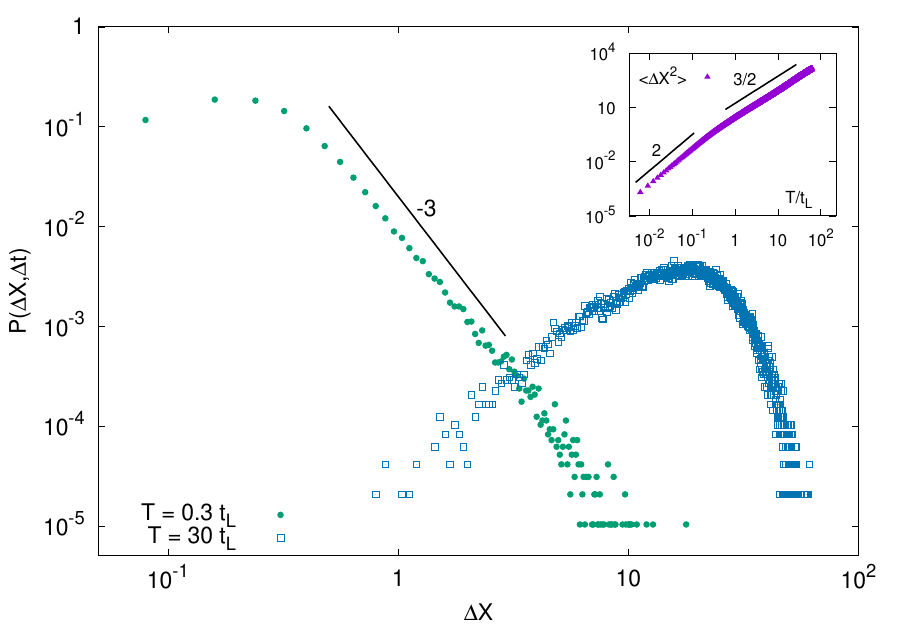}
  \caption{Absolute dispersion of droplets in the ageing regime at $\phi=77\%$.
    (Main panel). PDFs of droplet displacements (van Hove functions)
    at two time increments $T$, in the ageing regime; for
    the time increment in the ballistic regime the PDF shows, as expected, a $\Delta X^{-3}$ decay, consistent
    with the velocity PDF, Fig.~\ref{fig:pdfs_vel+acc}.
    (Inset). Mean square displacement, $\langle \Delta X^2 \rangle$ vs time increment; the solid lines highlight the
    ballistic and superdiffusive behaviours.}
\label{fig:pdfs+MSD}
\end{center}
\end{figure}

\begin{figure}
\begin{center}
  \advance\leftskip-0.55cm
  \includegraphics[scale=0.7]{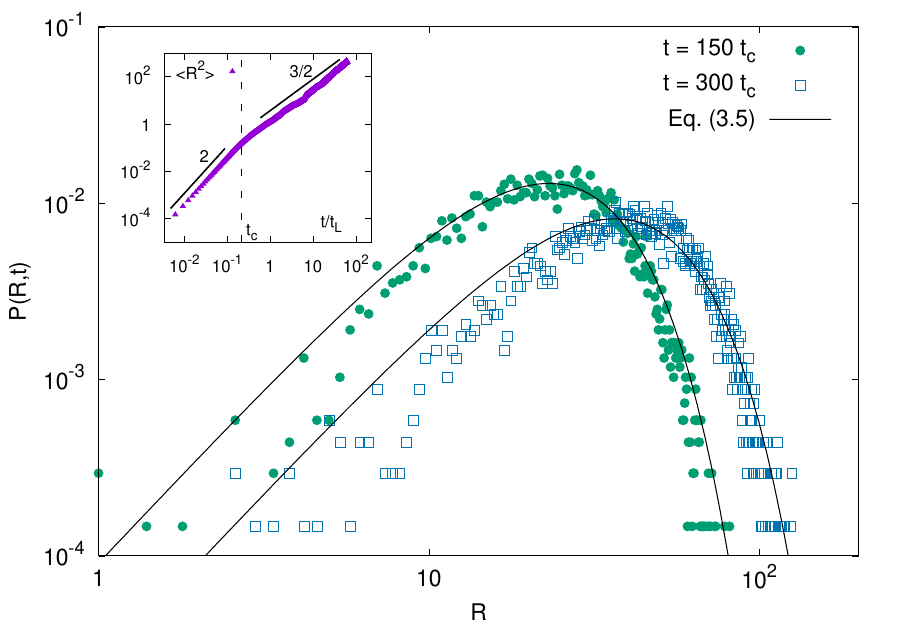}
  \caption{(Main panel). PDFs of pair separations, $P(R,t)$
    at two instants of time in the super-diffusive regime, $t_{1,2}>t_c$;
    the solid lines are the theoretical prediction, Eq.~(\ref{eq:pdftheo}), with parameters $\mathcal{A}=1.7$,
    $c \approx 0.25 t_c^{-1}$. (Inset). Mean square separation as a function of time: a ballistic regime,
    $\langle R^2 \rangle \sim t^2$, is followed by a super-diffusive one, $\langle R^2 \rangle \sim t^{4/3}$, the crossover
    between the two regimes occurring around $t \approx t_c = 12 d \eta/\gamma$, indicated with the dashed line.}
\label{fig:relative_andrea}
\end{center}
\end{figure}

In the ageing regime at the largest volume fraction, $\phi=77\%$, the
  MSD goes as $\langle \Delta X^2 \rangle \sim T^2$ for short times, signalling a ballistic regime, followed by
  a super-diffusive regime $\langle \Delta X^2 \rangle  \sim T^{3/2}$ (see inset of Fig.~\ref{fig:pdfs+MSD}).
The short time ballistic regime is consistent with a theoretical prediction based on
the superposition of randomly distributed elastic dipoles
(following structural micro-collapses)~\citep{BouchaudEPJE2001} and with results from experiments with colloidal
gels~\citep{CipellettiPRL2000} and foams~\citep{GiavazziJPCM2021}.
The scaling $\langle \Delta X^2 \rangle  \sim T^{3/2}$ is, instead, slightly steeper
than the experimentally measured $\sim T^{1.2}$~\citep{GiavazziJPCM2021}.\\
The ballistic regime $\langle \Delta X^2 \rangle  \sim T^2$ entails a power law tail of the
PDF of separations for short times, $P(\Delta X) \sim \Delta X^{-3}$, corresponding to the self part of the van Hove distribution~\citep{Hansen},
as reported in Fig.~\ref{fig:pdfs+MSD}~\citep{CipellettiFD2003,GiavazziJPCM2021}.
This observation finds correspondence, as one could expect, in the PDF of the droplet velocities,
shown in Fig.~\ref{fig:pdfs_vel+acc}.\\
The study of pair dispersion, reported in Fig.~\ref{fig:relative_andrea},
evidences, for the mean square pair separation (in the inset),
a ballistic regime, $\langle R^2 \rangle \sim t^2$, followed by a superdiffusive behaviour,
$\langle R^2 \rangle \sim t^{4/3}$.
The persistence of ballistic motion is expected to match the decorrelation of
trajectories following a plastic events, therefore the crossover time, $t_c$, can be approximately estimated
as the time taken by a droplet to travel over the typical size of a rearrangement, $\xi \approx 2 d $ (which is
an intrinsic scale for correlation lengths in soft glassy materials~\citep{Goyon2008,Dollet2015}); since
the characteristic velocity is $v_c \sim \frac{\Gamma}{6\eta}$ (see Fig.~\ref{fig:pdf_vel} and discussion thereof),
we get $t_c \sim \frac{12 d \eta}{\Gamma} \approx 7 \cdot 10^3$,
indicated in the inset of Fig.~\ref{fig:relative_andrea} with a dashed line.\\
%\begin{figure}[htbp]
By analogy with the Richardson's description of
turbulent diffusion~\citep{richardson1926,falkovich2001,boffetta2002}, we propose a phenomenological approach to derive the full pair separation PDF in the superdiffusive regime.
We assume that the such PDF evolves according
to a generalized diffusion equation, with a scale-dependent effective diffusivity, $D_{\text{eff}}$ which,
dimensionally, should be proportional to $\frac{d \langle R^2 \rangle}{dt}$. Since $\langle R^2 \rangle \sim t^{4/3}$ 
(and, consequently, $t \sim R^{3/2}$), we have
\begin{equation}
  D_{\text{eff}} \propto \frac{d \langle R^2 \rangle}{dt} \sim t^{1/3} \quad \Rightarrow \quad D_{\text{eff}} \propto R^{1/2}
  \equiv c R^{1/2}.
\end{equation}
The diffusion equation, thus, reads
\begin{equation}\label{eq:diffPDF}
\partial_t P(R,t) = \partial_{R} \left(cR^{1/2} \partial_{R} P(R,t)\right),
\end{equation}
that admits as solution (with the condition of unit area at all times) the following non-Gaussian distribution 
\begin{equation}\label{eq:pdftheo}
P(R,t) = \mathcal{A}\frac{2R^2}{27c^2 t^2}\exp\left(-\frac{4 R^{3/2}}{9ct}\right).
\end{equation}
The PDFs of pair separations measured at two instants of time in the superdiffusive regime, $t_1 \approx 150 t_c$ and
$t_2  \approx 300 t_c$, are shown in Fig.~\ref{fig:relative_andrea} together with the prediction of Eq.~(\ref{eq:pdftheo}),
with fitting parameter $c = 0.25 t_c^{-1}$, plotted as solid lines. The agreement obtained between
theory and numerics is quite remarkable.

\section{Conclusions}\label{sec:conclusions}
We presented results on the statistics of droplet velocities, accelerations and of droplet absolute
and relative dispersion in stabilized emulsions at various volume fractions, from semi-diluted to highly concentrated systems.
We employed a recently developed method for {\it in silico} emulsification of binary immiscible liquid mixtured with
high volume fractions of the dispersed phase, equipped with a novel tracking algorithm which allowed us to study
the emulsion physics at the droplet-resolved scale from a Lagrangian viewpoint, across various concentrations, from the semi-dilute to
jammed regimes.
Our results highlighted how the elastic properties and the plastic microdynamics of densely packed ensembles of droplets,
in close contact, 
are responsible of the non-Gaussian character of the droplet acceleration and, more moderately, of the velocity statistics.
We further investigated the single droplet diffusion in terms of both the mean square displacement and the self-part of the van Hove
distribution functions, finding that, while in the semi-dilute stirred case a ballistic-to-diffusive crossover is observed, in the
highly concentrated case a super-diffusive behaviour seems to emerge. Super-diffusion characterizes also the ageing regime, where
agreement is found with previous theoretical and experimental results.
Further investigations will focus on the dispersion properties on larger systems and for longer observation times, as well as
on the relation of the droplet Lagrangian properties with the stress distribution across the system.
In perspective, we foresee to extend the reach of the present work to extreme conditions of volume fractions and forcing amplitudes,
whereby the emulsion tends to loose stability and to undergo a catastrophic phase inversion. In this limit, too, the Lagrangian approach
is of invaluable utility.
Overall, our approach suggested a bridge between classical tools for Lagrangian high Reynolds number flows and 
complex fluid rheology, which paves the way to the inspection of unexplored aspects of the physics of soft materials.

\section*{Acknowledgements}
We are thankul to Chao Sun and Lei Yi for useful discussions and to Prasad Perlekar for a fruitful interaction at the
initial stage of the work. Numerical simulations were performed thanks to granted PRACE projects (ID: 2018184340 \& 2019204899)
along with CINECA and BSC for access to their HPC systems. This work was partially sponsored by NWO domain
Science for the use of supercomputer facilities.

\end{document}